\documentclass[journal=aamick,manuscript=article]{achemso}

\usepackage[version=3]{mhchem} % Formula subscripts using \ce{}
\usepackage[T1]{fontenc}       % Use modern font encodings
\usepackage{amsmath}
\usepackage{times}
\usepackage{textcomp}
\newcommand{\degree}{\ensuremath{^\circ}}
\newcommand{\mob}{cm\ensuremath{^2}/V$\cdot$s }
\newcommand{\om}{$\Omega\mu$m}
\newcommand{\mum}{$\mu$m }

\newcommand{\cmi}{cm\ensuremath{^{-1}}}

\newcommand{\ms}[1]{$_{\text{#1}}$}
\newcommand{\myref}[1]{Fig.~\ref{#1}}
\newcommand{\myrefEQ}[1]{Eq.~(\ref{#1})}

\author{Krishna Bharadwaj B}
\author{Rudra Pratap}
\author{Srinivasan Raghavan}
\email{sxraghavan@gmail.com}
\phone{}
\fax{}
\affiliation[Unknown University]
{Center For Nanoscience and Engineering, Indian Insitute of Science, India}

\title[An \textsf{achemso} demo]
  { Making Consistent Contacts to Graphene: Effect of Architecture and Growth Induced Defects}

\begin{document}

\begin{abstract}
  
	The effect of contact architecture, graphene defect density and metal-semiconductor work function difference on resistivity of metal-graphene contacts have been investigated. An architecture with metal on the bottom of graphene is found to yield resistivities that are lower, by a factor of 4, and most consistent as compared to metal on top of graphene. Growth defects in graphene film were found to further reduce resistivity by a factor of 2. Using a combination of method and metal used, the contact resistivity of graphene has been decreased by a factor of 10 to 1200 $\pm$ 250 \om{} using Palladium as the contact metal. While the improved consistency is due to the metal being able to contact uncontaminanted graphene in the metal on the bottom architecture, lower contact resistivities observed on defective graphene with the same metal is attributed to the increased number of modes of quantum transport in the channel.
	
\end{abstract}

%%%%%%%%%%%%%%%%%%%%%%%%%%%%%%%%%%%%%%%%%%%%%%%%%%%%%%%%%%%%%%%%%%%%%
%% Start the main part of the manuscript here.
%%%%%%%%%%%%%%%%%%%%%%%%%%%%%%%%%%%%%%%%%%%%%%%%%%%%%%%%%%%%%%%%%%%%%

\section{Introduction}
	In graphene devices, the metal-graphene contact resistivity  is a limiting  factor in various applications\cite{Balci2012,Li2013}. Particularly  in the case of radio frequency  devices and ultra low power sensor applications, ohmic contacts with resistivity in the range of 0.2-1 k\om{} are desirable\cite{Hong2015}.  Contact resistivity values as low as 200-500 \om{} have been reported on exfoliated graphene and devices of small dimensions \texttildelow100 nm). However, on the technologically relevant Chemical Vapor Deposition (CVD) grown graphene, the reported values are comparatively higher (0.5-10$^4$ \om\cite{Venugopal2010,Li2013,Song2014,Nagashio2009,Ma2014,Watanabe2012,Zhong2014}). Several methods have been employed to reduce the contact resistivity in CVD graphene devices. These include the uses of a sacrificial Al layer\cite{Hsu2011}, double contacted geometry\cite{franklin2012Double}, UV-Ozone treatment\cite{Li2013}, Ni catalyzed etching\cite{Wei2014} and localized plasma treatment/argon bombardment\cite{Choi2011}. Absolute values as low as  \texttildelow500 \om{} have been reported. However, from a device engineering perspective it is not only the absolute value of contact resistivity that is important but also the consistency in resistivity values achieved by a given method of contact formation. To illustrate the current state, a summary of the literature on the contact resistivity of Au-contacted graphene (Au is the most commonly used contact) on Si/SiO2  and the scatter in the reported values are shown in the Fig. S1 of supplementary material. Large scatter in the measured values of contact resistivity, as seen from the plot, is a common feature of measurements on graphene devices. For instance, a scatter of up 10$^4$(!) between reports using the same metal-graphene combination\cite{Venugopal2010,Li2013,Song2014,Nagashio2009,Ma2014,Watanabe2012,Zhong2014} as well as between devices in the same report have been observed.

\begin{figure}
		\centering
		\includegraphics[width=8.4cm,keepaspectratio]{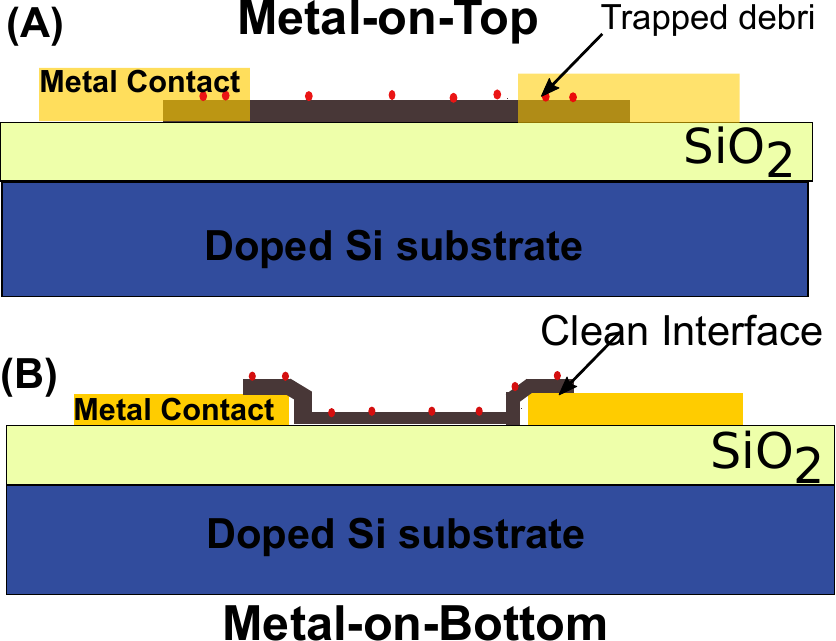}
		\caption{(A) MOTA or Metal-on-top architecture. (B) Metal-on-bottom architecture. The detailed process flow and figures are included in the supplementary material. The MOBA process flow precludes the possibility of there being process debri between the metal and graphene.}
		\label{fig:chap5-CR-deviceArchitecture}
	\end{figure}

	The contact resistivities reported in the literature are typically measured by using what we refer to in this letter as metal-on-top architecture (MOTA). \myref{fig:chap5-CR-deviceArchitecture}.(A) shows MOTA device schematics. This architecture has its origins in the fact that graphene research started with micron sized exfoliated monolayer flakes on a SiO$_2$/Si substrate whose physical location had to be optically identified prior to further lithographic processing\cite{Geim2009}. The contacts are then made on top of these flakes using electron beam assisted lithographic processes. This fabrication flow is susceptible to there being trapped photo-resist debris between the graphene and the metal leading to spurious contacts\cite{Wei2014,H2008} thereby leading to large variations (10$^3$ \om{} \cite{Venugopal2010}) in the observed contact resistivity. With the advent of large area monolayer graphene grown by CVD\cite{Li2009}, the constraint of having to first identify location is removed, allowing one to explore newer architectures such as the metal-on-bottom contact architecture (MOBA) contact topology used in the current study (\myref{fig:chap5-CR-deviceArchitecture}.(B)). We show that by using MOBA, consistent contacts can be obtained with resistivities that are lower, than those obtained on MOTA, by a factor of 4. 

	This effort at obtaining consistent contacts was initiated to ensure that a study on the effect of growth defects on the contact resistivity of CVD graphene is statistically relevant. Their effect on contact resistivity has been largely ignored\cite{Araujo2012}. Though recent reports on artificially created defects using ion bombardment methods and nano-particle based etching processes are available\cite{Wei2014}, they either require complicated processing steps or lack precise control on creating defects\cite{Li2013}. We show that in addition to reducing the metal-semiconductor barrier height, the growth induced defect density of the graphene film also helps to reduce contact resistivity. A reduction in contact resistivity by a factor of 2 is observed in this study for three of the most commonly used metals Au, Pt and Pd.  Thus, by using a combination of metal and method, a reduction in resistivity by a factor of 10 to values as low as 1200 $\pm$ 250 \om{} is demonstrated within a controlled set of experiments. As shown in the supplementary material, this is one of the best combinations yet reported.

\section{Experimental methods}
	
	Graphene was grown on Cu foil using CVD in two different conditions as described in the supplementary material. During graphene growth,  two different defect densities were obtained by controlling the source flow conditions as described by \citet{Tsen2012}.  Post CVD, PMMA (950 A4) was spin coated on the graphene covered Cu substrate and underlying Cu was etched using ammonium per sulphate solution. Following Cu etch, the PMMA-graphene composite was further processed by two different methods to yield the MOBA and MOTA architectures as described below. 

	For device fabrication a heavily p-doped Si wafer with 300 nm SiO$_{2}$ capping was used as the initial substrate. For the MOBA, metal electrodes were first patterned on the substrate using standard optical lithographic lift-off processes as shown in supplementary material Fig. S3. A mild oxygen plasma treatment was performed to remove photo resist residues and provide a clean surface to contact graphene. The graphene polymer composite was transferred on top of these patterned metal pads to yield bottom contacted graphene. Since, the metal surface on which graphene is transferred has never been covered by polymer at any stage in processing, this contact architecture provides cleaner metal-graphene interface. After drying at room temperature for 8 hrs, a soft bake on a hot plate at 180\degree C in air was done to remove trapped water and improve adhesion. The PMMA support layer was finally removed with an overnight acetone etch thereby yielding MOBA contacts. In the MOTA process, graphene was first directly transferred on top of a clean SiO$_{2}$ surface using the PMMA support layer as described previously. Following PMMA removal using acetone, metal pads and graphene channel were patterned using standard lithographic procedures as described in the supplementary material.

\section{Results and Discussion}
	
\begin{figure}
		\centering
		\includegraphics[width=8.46cm,keepaspectratio]{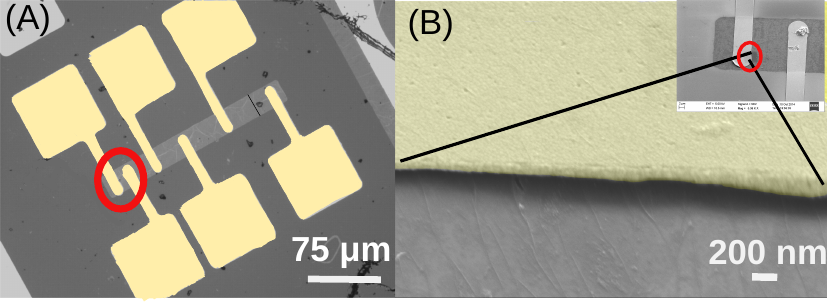}
		\caption{Device Representation:(A) Typical TLM structure fabricated using the MOBA. 25 \mum wide channels are fabricated with lengths varying from 10 \mum to 50 \mum. (B) A magnified image of the metal-SiO$_{2}$ interface (see text for discussions). No tears are visible.}
		\label{fig:chap5-CR-deviceSEM}
	\end{figure}

	Transfer length measurement  structures are typically  used to measure the contact resistivity of metal semiconductor junctions\cite{Dieter2006}. Graphene channels of varying lengths were fabricated as  shown in \myref{fig:chap5-CR-deviceSEM}.(A). The channels were made 25 \mum wide to average the statistical variations in the number of grains (usual grain sizes were 2-5 \mum as shown in the supplementary material) and the device lengths were varied from 10 \mum to 50 \mum in 5 steps. Since, in MOBA, the graphene is transferred onto the metal contact pads, which are usually 100 nm thick, it is important to examine the region, (see encircled in red (online only) in \myref{fig:chap5-CR-deviceSEM} in which graphene monolayer comes off the metal surface contacts onto the wafer to ensure that there are not tears there. SEM image of this region is shown in \myref{fig:chap5-CR-deviceSEM}.(B). In the inset, graphene is seen in dark contrast with the bright metal layer at the bottom. Due to the thick support layer of PMMA used during transfer, graphene remains intact and is defect free in this region.  Also, it is important to note that graphene contacts the SiO$_2$ layer within 200 nm from the metal side wall making the suspended graphene region negligible, in comparison with the large channel lengths considered.

\begin{figure}[h]
		\centering
		\includegraphics[width=8.46cm,keepaspectratio]{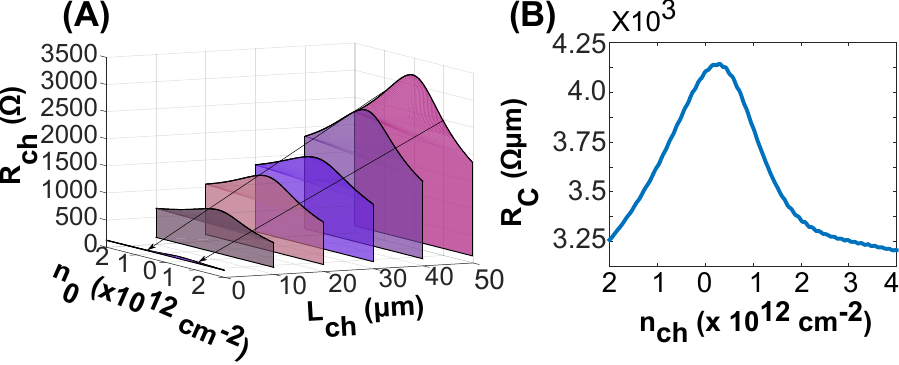}
		\caption{ Carrier density adjusted TLM: (A) Channel resistance R\ms{ch} of gFETs (made using Au contact pads) having different channel lengths L\ms{ch} are plotted against gate dependent carrier concentration n$_{0}$. Straight line fits made to resistances at the same carrier concentration are used to extract the contact resistance R\ms{C} (in $\Omega$, which is later multiplied by width to obtain resistivity) by using the Y-intercept. (B) The contact resistivity so extracted.}
		\label{fig:chap5-CR-IDVG}	
	\end{figure}

	With  the heavily doped Si as the bottom gate, each  pair of electrodes in the TLM structure serves as a source-drain contact to the graphene field effect transistors (gFET) giving rise to a gated TLM structure. Due to the nonexistent band gap and a continuously gate tunable channel carrier concentration (and hence the Fermi energy, E$_f$ ), a gate voltage dependent contact resistivity is expected\cite{Xia2011}, though instances of gate-independent contact resistivities have also been reported\cite{Russo2010}. Two terminal channel resistivity vs.\ gate voltage (V\ms{G}) characteristics of the individual gFETs with varying lengths that constitute the TLM  structure are plotted in \myref{fig:chap5-CR-IDVG}.(A), with the longest one having the highest resistance.  All the devices showed clear p-type behavior (indicated by the positive charge neutrality point (CNP) and the corresponding gate voltage V\ms{CNP}) as is usually the case in CVD grown graphene due to unintentional doping\cite{Venugopal2011}. The CNP and corresponding gate voltage V\ms{CNP}  of all the devices were close but not coincident and were in the range of 15-20 V (on a 300 nm oxide, 15 V translates to 0.5 MV/cm).  The uniform and monotonic shift in CNP at different channel lengths (shown in the Fig S3 of supplementary material) clearly indicates a uniform doping density in graphene, a hallmark of CVD grown films. To account for the non-coinciding V\ms{CNP} in the different devices, the contact resistivity of the entire TLM structure was extracted  as a function of the carrier concentration (n\ms{ch}) derived using \myrefEQ{eq:chap5-eq1}, where C\ms{ox} is the gate oxide capacitance, n\ms{0} is  the intrinsic carrier concentration and n\ms{G} is the gate dependent carrier concentration, as opposed to the normally used gate voltage. \myref{fig:chap5-CR-IDVG}.(A) stacks the R-n\ms{ch} plots of all the devices in the TLM structure. The black lines connect the resistivity values obtained at the same n\ms{ch} on various devices and are extrapolated to obtain the contact resistivity. The extracted contact resistivity in \om{} is shown in the \myref{fig:chap5-CR-IDVG}.(B).
	
	\begin{subequations}
		\begin{align}
			n_{\text{ch}} = n_0 + n_G \\
			n_{G} = (V_G - V_{CNP})*C_{ox}/q \\	
			\rho_{ch} = \frac{1}{\sqrt{{n_{0}}^2 + {n_{G}}^{2}} * \mu{}}		
		\end{align}
			\label{eq:chap5-eq1}
		\end{subequations}	
	
\begin{figure}[h]
		\centering
		\includegraphics[width=8.46cm,keepaspectratio]{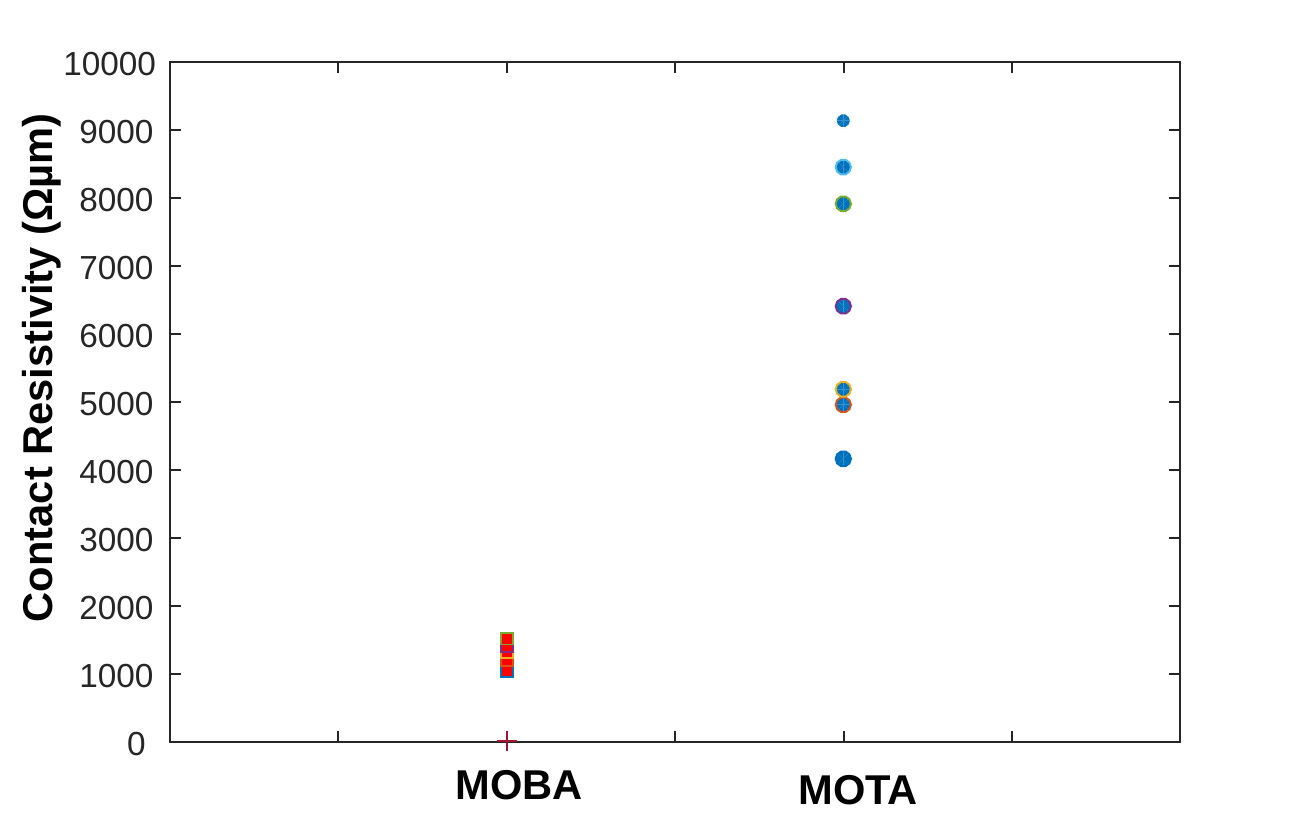}
		\caption{Effect of MOBA vs.\ MOTA on contact resistivity measured using Au contact pads and at zero gate bias. MOBA not only has lower contact resistivity, but also better consistency in the observed values.}
		\label{fig:chap5-CR-topBottom}
	\end{figure}

	 Figure ~\ref{fig:chap5-CR-topBottom} compares the contact resistivity, when Au is used as the contact metal, extracted at zero gate voltage from the TLM structures fabricated using MOTA and MOBA. The values obtained using similar graphene (from the same growth condition) on MOBA was not only four folds lower in the absolute magnitude, but also in the standard deviation. This indicates that, the `graphene-last' fabrication process, used in MOBA, significantly reduces the contamination and lithographic residues at the metal-graphene interface, thereby enabling better consistency in the observed values.
	 
	 While the lowest reported contact resistivity in ex-foliated graphene is 128 \om{} \cite{Malec2011}, the typical reported values using CVD grown graphene with Au is 790$\pm$300 \om, both using MOTA\cite{Venugopal2010}. The lowest absolute contact resistivity obtained in this work is 1200$\pm$250 \om{} and it compares reasonably well with the usually reported values ((1000-10000) \om) in the literature\cite{Nagashio2009,Venugopal2012}. However, a low overall contact resistivity was not the objective of this work. Rather it was the comparison between bottom and top contacts obtained on the same platform and the effect of graphene defects. Since, in the previous section, the bottom contacted devices showed better consistency in the observed values, MOBA architecture was used to observe the effect of defect density and metal work function difference on the contact resistivity as described below.

\begin{figure}
		\includegraphics[width=8.46cm,keepaspectratio]{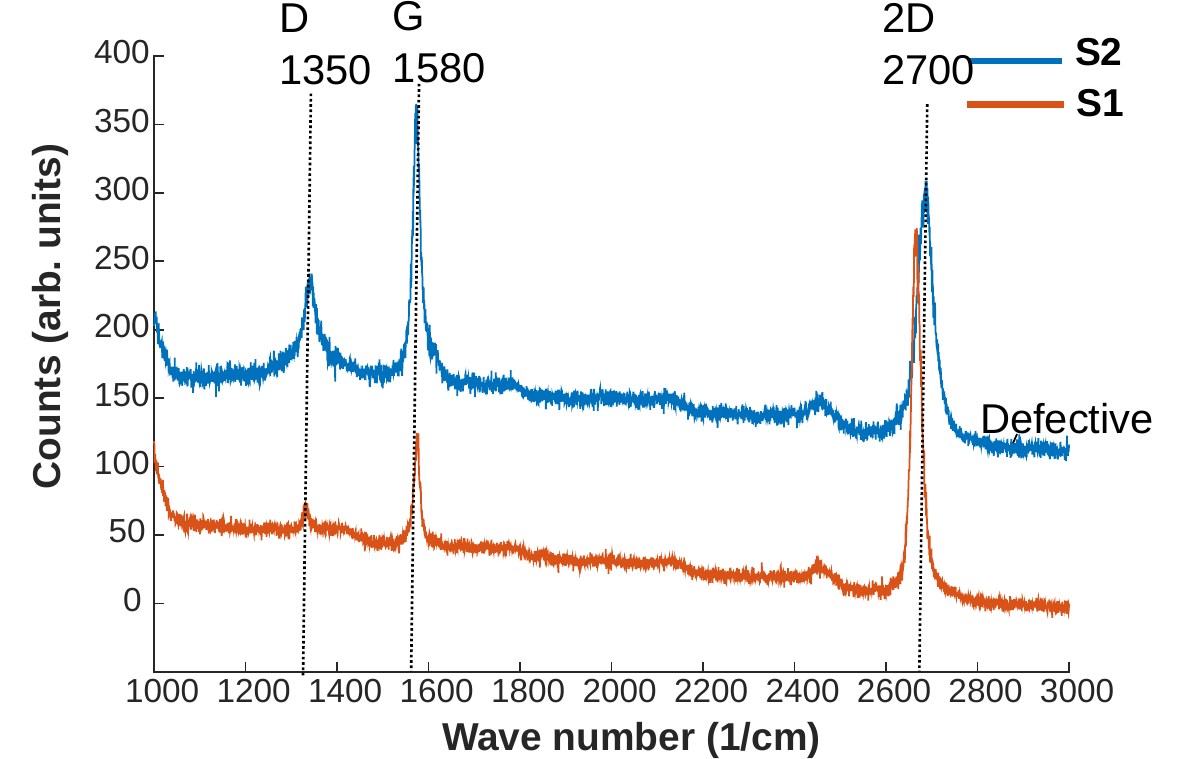}
		\caption{Raman spectrum obtained on samples S\ms{1} and S\ms{2}. The 'D' peak intensity, which can be used to quantitatively determine the defect density indicates an order of magnitude higher defect density in S\ms{2} than in S\ms{1}.}
		\label{fig:chap5-CR-raman}
	\end{figure}

\begin{figure}
		\centering
		\includegraphics[width=8.46cm,keepaspectratio]{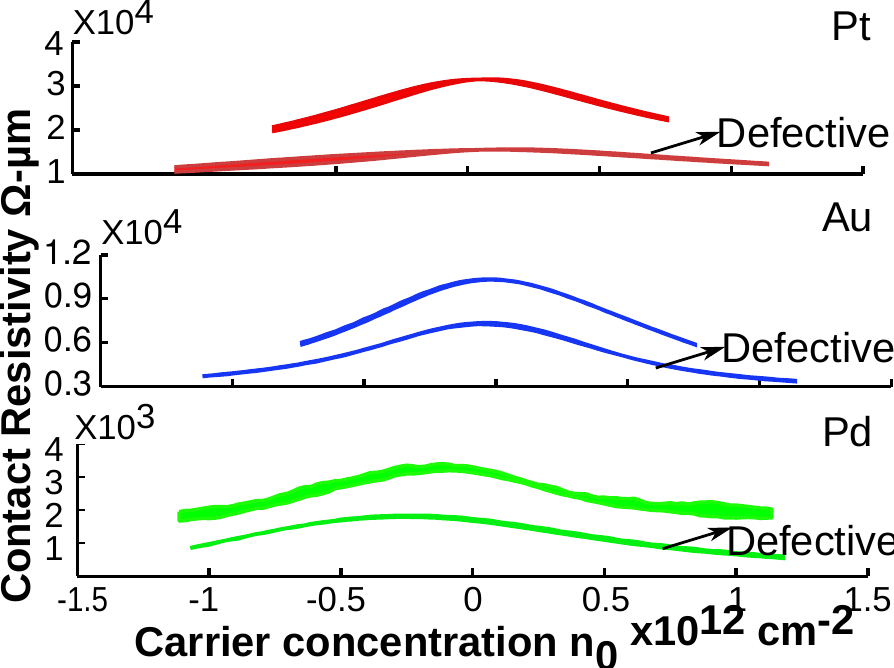}
		\caption{Effect of metal work function and graphene defect density on contact resistivity. The contact resistivity obtained at carrier concentration varying from -1.5 - 1.5 x 10$^12$ cm$^{-2}$ are plotted for three metals namely Pt, Au and Pd in decreasing order of their metal work functions. The values with the standard deviation are displayed as bands in each plot. For each metal, the contact resistivity of graphene with two different defect densities are plotted. From the figure it is clearly seen that under the same metal, the graphene with higher defect density has a lower resistivity. Moreover, under this architecture, the resistivity shows a linear relation with the metal work function.  }
		\label{fig:chap5-CR-cRMetals}
	\end{figure}

	The effect of graphene growth defect density on the contact resistivity has been a hitherto unexplored area despite its obvious practical importance in devices. Graphene films having different defect densities were grown by controlling the source flow conditions as described in the supplementary material, and the Raman spectroscopic studies obtained on the defective sites on two different samples used (S$_{1}$ and S$_{2}$) are shown in \myref{fig:chap5-CR-raman}. The 'D' peak at 1350 \cmi('D' peak) qualitatively indicates the number of defects in the graphene sample\cite{Ferrari2006}. A simple comparison of the peak intensities indicate that S$_{2}$ has a higher defect density. From the expression developed by \citet{Cancado2011}, it is seen that S$_{2}$ has an order of magnitude higher defect density than S\ms{1}. The three most commonly used metals, namely Pt, Au, and Pd with work functions 5.65 eV, 5.40 eV, and 5.22 eV respectively were chosen to fabricate devices using the two kinds of graphene. Figures~\ref{fig:chap5-CR-cRMetals} shows the contact resistivity extracted from TLM measurements. It can be seen that irrespective of the metal used, contact resistivity is lower when the graphene film is defective by a factor of 2. In contrast, the sheet resistance increases with defect density, making the inverse dependence of contact resistivity on defect density, an intriguing observation. From the I\ms{D}-V\ms{G} of the individual devices, the field effect mobility was extracted and the residual carrier concentration was calculated using equation \myrefEQ{eq:chap5-eq1} (details on the extaction and fit are presented in the supplementary material). Comparing the values obtained using the constant mobility fit, it was seen that the sample S\ms{2} with higher defects has a mobility of 1200 \mob and a intrinsic carrier concentration of 1E13 cm$^{-2}$, while sample S\ms{1} had a mobility of 16310 \mob and a carrier concentration of 2.5E12 cm$^{-2}$. 
			
	The contact conductivity (G$_{c}$) of graphene has been modeled by ~\citet{Xia2011} as follows,
	\begin{subequations}
	\begin{flalign}
		G_{c} = (4e^2/h)TM_{total} \\\label{eq:chap5-eq2-a}
		M =(\Delta E_f/\pi \hbar V_f)W  \\\label{eq:chap5-eq2-b}
		M_{total} = M_{VG} + M_{CNP} \\\label{eq:chap5-eq2-c}
		M_{CNP} =(E_{ex}/\pi \hbar V_f)W\\\label{eq:chap5-eq2-d}
		E_{ex} = \hbar v_f \sqrt{\pi n_0}
	\end{flalign}
	\label{eq:chap5-eq2}
	\end{subequations}
	 where M is the quantum number of modes on graphene, T is the transmission probability from metal to graphene in the bulk, e is the electronic charge, E\ms{F} is the Fermi level at a given gate voltage and E\ms{ex} is the Fermi level of intrinsic graphene, V\ms{f} is the Fermi velocity, $\hbar$ is the reduced planks constant, and W is the channel width. The quantum number of modes in graphene M is related to $\Delta E_f$ as described in ~\myrefEQ{eq:chap5-eq2-b}, where $\Delta E_f$ is the energy difference between the actual Fermi level in the system ($E_F$) and charge nuetrality point energy (E\ms{CNP}). In intrinsic graphene, the carrier concentration drops to zero at $E_{F} = 0$ due to vanishing density of states, and the conductance drops to the minimum conductance level ($\sigma = 2e^{2}/\hbar$). However, CVD grown graphene gets unintentionally doped and is usually defective leading to higher conductance at the CNP. This increased conductance at the CNP is due to additional number of conductance modes made available due to the defects in the sample. Hence the total number of modes in the graphene channel is $M_{total} = M_{VG} + M_{CNP}$, where $M_{VG}$ is the number of modes at a particular $E_{F}$, obtained as described before and $M_{CNP}$ is the number of modes in the channel due to the presence of defects. If the carrier concentration at CNP (n\ms{0}) is known, the excess number of modes at CNP can be calculated as mentioned in equation set ~\myrefEQ{eq:chap5-eq2}.

\begin{figure}[h]
		\centering
		\includegraphics[width=8.46cm,keepaspectratio]{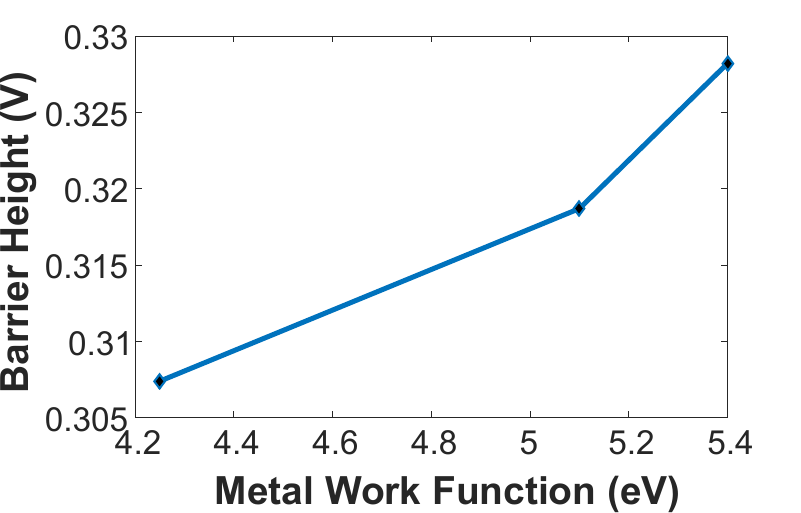}
		\caption{Plot of barrier height vs.\ metal work function. Linear relation between the two indicates the thermionic emission being the prominent interface conductivity and thus, supporting the evidence that the contact resistivity is proportional to the work function difference}
		\label{fig7:chap5-CR--wf}
	\end{figure}

	Using the residual carrier concentration extracted previously on both the samples and using the ~\myrefEQ{eq:chap5-eq2}, the E$_{ex}$ on the defective graphene was higher than the other by 0.3 meV, resulting in $M_{CNP}(B) = 2 * M_{CNP}(A) $. The contact resistivity ratio of the two samples with Pt at the CNP, as seen from the ~\myref{fig:chap5-CR-cRMetals}, was only slightly higher than 2 indicating that the change in number of modes(M) under the defective graphene could be the dominant reason for the reduction in the contact resistivity in defective graphene. However, it should be noted that introducing defects to reduce contact resistivity might adversely affect the device performance as the uniform distribution of defects in graphene has a degrading effect on the charge mobilities\cite{hwang2010correlating}. While such a method of reducing the contact resistivity would benefit sensing applications, whose sensitivity improves with defect density the film, high speed logic applications might be affected due to mobility degradation. Hence, it is important to choose the right defect density to be introduced in the material such that a right balance obtained between the channel resistance and the contact resistivity and thereby satisfying the application requirements. Since such an observation is not unique to graphene, but also many other 2D materials, these results hence offer an interesting route to further reduce contact resistivity to atomically-thin devices by selective defect engineering. Also, in the quest to find the most suitable material for graphene contacts, several metals have been previously explored in the literature\cite{Ma2014,nagashio2010,Joshua2011,Hsu2011}. There are considerable number of contradicting reports and there is no final verdict on the best suited material for a given architecture\cite{Chaves2014,Watanabe2012}. As the effect of defects in graphene on the contact resistivity was largely ignored previously, it is possible that this might be the reason for the discrepancy. Figures ~\ref{fig:chap5-CR-cRMetals}  plots the contact resistivity vs.\ carrier concentration of graphene with various metals using MOBA. At the same carrier concentration, with the graphene being the same across different metals, Pd had the lowest contact resistivity among the three, followed by Au and Pt. Using a Schottky contact model between graphene and metal, the barrier height can be extracted as described by \citet{Dieter2006}. The linear relation between the barrier height (extracted from the Rc) and the metal work function shown in \myref{fig7:chap5-CR--wf} indicates that thermionic emission is the dominating transport at the interface\cite{sze2006physics}.  
		
	\section{Summary}	
		In summary, we have extracted the contact resistivity of graphene using a minimal damage graphene-last fabrication process leading to a bottom contact architecture. This architechture results in very consistent contacts with low resistivities. The higher consistency allows a statistically relevant study of other effects such as defects on contact resistivity. It was seen that on all three commonly used materials namely Pt, Pd and Au, growth defects reduce contact resistivity by a factor of 2. Pt with the highest work function had the highest contact resistivity and Pd the lowest

\bibliography{paper_3.bib}

%%%%%%%%%%%%%%%%%%%%%%%%%%%%%%%%%%%%%%%%%%%%%%%%%%%%%%%%%%%%%%%%%%%%%
%% The appropriate \bibliography command should be placed here.
%% Notice that the class file automatically sets \bibliographystyle
%% and also names the section correctly.
%%%%%%%%%%%%%%%%%%%%%%%%%%%%%%%%%%%%%%%%%%%%%%%%%%%%%%%%%%%%%%%%%%%%%
%\bibliography{achemso-demo}

\end{document}